# Stabilization and Controlled Association of Inorganic Nanoparticles using Block Copolymers

**K. Yokota, M. Morvan, J.-F. Berret[@]**
Complex Fluids Laboratory, UMR CNRS - Rhodia n°166,
Cranbury Research Center Rhodia 259 Prospect Plains Road, Cranbury NJ 08512 USA

**and J. Oberdisse**
Groupe de Dynamique des Phases Condensées, UMR CNRS – Université de
Montpellier II n° 5581, F-34095 Montpellier France

**Abstract :**

We report on the structural properties of mixed aggregates made from rare-earth inorganic nanoparticles (radius 20 Å) and polyelectrolyte-neutral block copolymers in aqueous solutions. Using scattering experiments and Monte Carlo simulations, we show that these mixed aggregates have a hierarchical core-shell microstructure. The core is made of densely packed nanoparticles and it is surrounded by a corona of neutral chains. This microstructure results from a process of controlled association and confers to the hybrid aggregates a remarkable colloidal stability.

Inorganic nanoparticles are currently used in a wide variety of material science applications such as catalysis, protective coating or opto-electronics. For applications, inorganic particles often require to be concentrated, dried or mixed in various environments. These transformations usually result in dramatic changes for the suspensions, for instance in a destabilization or an irreversible precipitation of the particles [1,2]. A broad range of techniques in chemistry and physical chemistry has been developed in the last years to stabilize inorganic nanoparticles [2-6], or to self-assemble them in a controlled manner [7,8]. Among these techniques, one can cite the adsorption of charged ligands or stabilizers on their surfaces [5], the layer-by-layer deposition of polyelectrolyte chains [9] or the surface-initiated polymerization resulting in a high-density polymer brushes [10]. In this letter we explore an alternative route for the stabilization of nanoparticles, based on the principle of complexation between oppositely charged species. The nanoparticles are complexed using asymmetric block copolymers, where one block is charged and of opposite charge to that of the particles and the second block is neutral.

It has been shown recently that polyelectrolyte-neutral copolymers can associate in aqueous solutions with oppositely charged surfactants and build stable "supermicellar" aggregates with core-shell structures [11-13]. The formation of the mixed aggregates occurs according to the mechanism of electrostatic self-assembly [14]. The core (of radius $\sim 100 - 200$ Å) is described as a complex coacervation micro-phase and it is made from surfactant micelles connected by the polyelectrolyte blocks. The corona (of thickness $\sim 250 - 500$ Å) is made by the neutral chains and insures the stability of the whole [15]. The electrostatic self-assembly observed with block copolymers [11,13,16-19], as well as the phase separation found more generally in polyelectrolyte-macroion mixtures [20-23] occur for charge ratios of the order of unity, with the possibility in some cases of redissolution at high charge ratios. Several theoretical models have been proposed to account for the phenomenon of complexation in synthetic [14,20,24,25] and biological [26,27] systems.

Here we show that the electrostatic self-assembly is appropriate to stabilize inorganic nanocolloids as well. We study the complexation between yttrium hydroxyacetate nanoparticles and poly(acrylic acid)-*b*-poly(acrylamide) block copolymers. The yttrium particles were chosen as a model system for inorganic colloids. From their structural and electrostatic properties they are similar to surfactant micelles. The yttrium nanoparticles were also considered because of their applications as precursors for ceramic and opto-electronic materials. Using scattering techniques and Monte Carlo simulations, we determine consistently the sizes, aggregation numbers and microstructure of the hybrid organic-inorganic complexes.

***Structural properties of the yttrium-based nanoparticles***. The synthesis of the yttrium hydroxyacetate nanoparticles is based on two chemical reactions, the dissolution of yttrium oxide by acetic acid at elevated temperature and the controlled reprecipitation of yttrium hydroxyacetate occurring on cooling, yielding the average composition $Y(OH)_{1.7}(CH_3COO)_{1.3}$ for the particles [28,29]. Here, we focus on the structural properties of these nanoparticles. Using light scattering, we have determined their weight averaged molecular weight $M_w^{nano} = 27\,000$ g mol$^{-1}$ and their hydrodynamic radius $R_H^{nano} \sim 20$ Å. Zeta potential measurements confirm that they are positively charged ($\zeta = + 45$ mV) and that the suspensions are primarily stabilized by surface charges. The stability of $Y(OH)_{1.7}(CH_3COO)_{1.3}$ suspensions at neutral pH is excellent over a period of several weeks. Small-angle x-ray scattering (SAXS) experiments performed on the yttrium sols at different concentrations have shown that the scattering cross-section can be factorized as [30,31] :

$$\frac{d\sigma}{d\Omega}(q,c) = n(c)\Delta\rho^2 V^2 S(q,c) F(q) \qquad (1)$$

In Eq. 1, $n(c)$ is the number density, $\Delta\rho$ the scattering contrast and $V$ the volume of the particles. $F(q)$ and $S(q,c)$ are the form and structure factors, respectively. The SAXS experiments were carried out at the Brookhaven National Laboratory on the X21 beam line, using an incident wavelength of 1.76 Å and a sample-to-detector distance of 1 meter, so as to cover the wave-vector range $0.008 - 0.5$ Å$^{-1}$. Fig. 1 displays $S(q,c)$ for c comprised between 0.8 and 25 wt. %. The decrease of the structure factors as $q \rightarrow 0$ indicates strong interparticle repulsions, which are attributed to steric and electrostatic interactions. In the inset, the low q-behavior of the scattering cross-section at $c = 0.4$ wt. % is shown as function of $q^2$ in semilogarithmic scale (Guinier representation). At this concentration, $S(q) \sim 1$ and the intensity describes the form factor of the nanoparticles. The linear dependence in the inset corresponds to a radius of gyration $R_G^{nano} = 16.5$ Å, a value in good agreement with that of the hydrodynamic radius.

***Complexation properties and stabilization***. The yttrium-based nanoparticles were complexed with poly(acrylic acid)-*b*-poly(acrylamide) block copolymers (polydispersity index 1.6) [28]. Poly(acrylic acid) is a weak polyelectrolyte which ionization state depends on the pH. All the experiments were conducted at pH 7 where $\sim 70$ % of monomers are negatively charged. In the following, the copolymer is referred to as PANa(69)-*b*-PAM(840), the two numbers in parenthesis being the degrees of polymerization of each block. These numbers correspond to molecular weights 5 000 and 60 000 g mol$^{-1}$, respectively. Polymer-nanoparticles complexes were obtained by mixing the pure solutions prepared at the same concentration c and pH. The mixing ratio X is defined as the volume of yttrium-based suspension relative to that of the polymer. With oppositely charged species, it is usual to describe the mixed solutions in terms of charge ratios. This is not possible here since the structural electrostatic charges borne by the particles is not known [28].

Fig. 2 shows the X-dependence of the hydrodynamic radius $R_H$ determined by dynamic light scattering experiments. We have checked that the concentration fluctuations from the mixed solutions are diffusive. The hydrodynamic radius is calculated according to the Stokes-Einstein relationship, $R_H = k_B T / 6\pi\eta_S D(\mathbf{c}\rightarrow 0)$ where $D(\mathbf{c})$ is the collective diffusion coefficient, $k_B$ is the Boltzmann constant, T the temperature (T = 298 K) and $\eta_0$ the solvent viscosity ($\eta_S = 0.89\times10^{-3}$ Pa s). Below the critical mixing ratio $X_C = 0.1$, $R_H$ is that of the single PANa(69)-$b$-PAM(840) chains ($R_H^{pol} = 79$ Å). Above $X_C$ and over three decades in X, the hydrodynamic radius is narrowly distributed and constant around 300 Å. $R_H$-values well above those of the individual components actually suggest the formation of mixed aggregates. For large X (X $\geq$ 10), the autocorrelation functions of the scattered light are described by a double exponential relaxation. The fast mode corresponds to the diffusion of single nanoparticles ($R_H^{nano} \sim 20$ Å) and the slow mode to that of the mixed aggregates mentioned previously. After a period of 6 months however, the solutions at large X as well as the nanosol become turbid. The solutions contain sub-micrometric clusters in addition to the complexes. These clusters result from the aggregation of the formerly unassociated nanoparticles. In the low X-range on the contrary (X < 1), the mixed solutions remain unchanged over the same period, suggesting that the nanoparticles have been stabilized through the association with the copolymers.

***Determination of the aggregation numbers***. With electrostatic self-assembly, there is usually a mixing ratio $X_P$ at which all the components present in solution react and form complexes [16,19]. $X_P$ corresponds roughly to a state where the cationic and anionic charges are in equal amounts [14]. Experimentally, it is the ratio where the number density of complexes is the largest, *i.e.* where the scattering intensity as function of X presents a maximum. From the X-dependence of the Rayleigh ratio monitored on the solutions of Fig. 2 (data not shown), we have found $X_P \sim 0.2$ [29]. In the following, we focus only on this composition. Fig. 3 shows results of static light scattering. The data are displayed in a Zimm representation, which consists in plotting the quantity $Kc/\mathcal{R}(q,\mathbf{c})$ as function of $q^2 + cste \times c$, where K denotes the scattering contrast, $\mathcal{R}(q,\mathbf{c})$ the Rayleigh ratio and $q = \dfrac{4\pi n}{\lambda}\sin(\theta/2)$ the scattering vector. In the above definitions, n is the refractive index, $\lambda$ the wavelength of the incident beam and $\theta$ the scattering angle. The scattering contrast $K = 4\pi^2 n^2 (dn/dc)^2/\mathcal{N}_A\lambda^4$ was determined by separate measurements of the refractive index increment $dn/dc$ ($\mathcal{N}_A$ is the Avogadro number). $Kc/\mathcal{R}(q,\mathbf{c})$-data are displayed in Fig. 3, together with fits using the classical expression for macromolecules and colloids [31] :

$$\frac{Kc}{\mathcal{R}(q,c)} = \frac{1}{M_{w,\mathrm{app}}}\left(1 + \frac{q^2 R_G^2}{3}\right) + 2A_2c \qquad (2)$$

In Eq. 2, the radius of gyration $R_G$, the weight-averaged molecular weight $M_{w,\mathrm{app}}$ and the second virial coefficient $A_2$ of the mixed nanoparticle-polymer aggregates are physical characteristics which are determined by fitting. One gets $R_G = 190 \pm 15$ Å $M_{w,\mathrm{app}} = 3.2\times10^6$ g mol$^{-1}$ and a virial coefficient of $10^{-6}$ cm$^{-3}$ g$^{-2}$ [29]. In Fig. 3, we verify in addition that dilution does not change the state of aggregation of the complexes [13,16,18,19]. For colloids resulting from a self-assembly process, the apparent molecular weight can be expressed as $M_{w,\text{app}} \approx \left( \overline{n^2}/\overline{n} \right) m_n^*$. Here, $\overline{n}$ ($\gg 1$) and $\overline{n^2}$ are the first and second moments of the distribution of aggregation numbers and $m_n^*$ is the molecular weight (number-average) of the elementary building blocks which constitute the colloid [30]. By choosing for building block a unit composed by one nanoparticle and two polymers (this proportion is actually fixed by $X_P$ [29]), the aggregation number appearing in $M_{w,\text{app}}$ coincides with the number of particles per aggregate, $n^{nano}$. $m_n^*$ and $\overline{n^{nano^2}}/\overline{n^{nano}}$ can then be estimated with the help of the molecular weights and polydispersities of the single components. For the solutions in Fig. 3, we find $\overline{n^{nano^2}}/\overline{n^{nano}} = 31$ and a number that is roughly twice this value for the polymers. As a comparison, in surfactant-polymer mixtures, aggregation numbers in the "supermicellar" aggregates were estimated of the order of hundreds, with typically one surfactant micelle per polymer. This suggests that the structural charge of the yttrium hydroxyacetate nanoparticles might be larger than for micelles.

***Internal structure of the colloidal complexes.*** Fig. 4 illustrates the SAXS intensities obtained at c = 1 wt. % and X varying between X = 0 and X = 10. The data were taken at the Brookhaven facility using the scattering configuration of Fig. 1. For the pure polymer solution (X = 0), the overall intensity is weak and it follows at high q a scaling law with an exponent −1.4 [30,31]. Data for X = ∞ represent the form factor of the nanoparticles, which low q-dependence was already discussed (Fig. 1). The entire q-range reveals now a bump around q ∼ 0.3 Å$^{-1}$ that we have ascribed to the internal structure of the yttrium hydroxyacetate nanoparticles. For the intermediate X-values, X = 0.2, 1 and X = 10, the scattering cross-section is dominated by a strong forward scattering. Note the similarities in the q-dependences between the data at X = 10 and X = ∞. This shows again that at high X there is a coexistence between unassociated nanoparticles and mixed aggregates.

In order to interpret the spectra of Fig. 4 quantitatively, we assume for the nanoparticle-

polymer complexes a core-shell structure similar to the one found with surfactant micelles

[13,15]. In this model, the core is a dense assembly of particles, which dominates the

scattering intensity in the wave-vector range of interest (q > 0.01 Å$^{-1}$). The corona is a diffuse

shell which is detected at much lower q, such as by light scattering. It was shown in Ref. [15]

that the scattering cross-section for such an assembly has an analytical form identical to Eq. 1.

It is the product of the form factor of the single particle with the function :

$$S(q, n^{nano}) = 1 + \frac{1}{n^{nano}} \sum_{i \neq k}^{n^{nano}} \frac{\sin[q(r_i - r_k)]}{q(r_i - r_k)} \quad (3)$$

$S(q, n^{nano})$ is derived from the Debye formula for models made of $n^{nano}$ identical scatterers

and located at the respective positions $r_i$ in the aggregate [31]. For simplicity we neglect here the interactions between complexes. Eq. 3 has been calculated by Monte Carlo simulations using a hard sphere interaction potential between particles and for different aggregation numbers and polydispersity indexes $s^{nano}$. For polydisperse cores, the asymptotic values of the function in Eq. 3 are : $S(q \rightarrow 0, n^{nano}, s^{nano}) = \overline{n^{nano^2}}/\overline{n^{nano}}$ and $S(q \rightarrow \infty, n^{nano}, s^{nano})$

= 1 [15]. In Fig. 5, we have reproduced the intensity obtained for the solution prepared at c =

1 wt. % and X = 0.2. In the range 0.15 − 0.5 Å$^{-1}$, the scattering intensity superimposes well with the form factor of the individual nanoparticles, confirming thus the validity of Eq. 1.

Similar results were found for mixing ratios between X = 0.1 and 10. In the inset of Fig. 5, the

experimental function $S(q, n^{nano}, s^{nano})$ is compared with the results of a simulation using

$\overline{n^{nano^2}}/\overline{n^{nano}} = 39$, $\overline{n^{nano}} = 26$ and $s^{nano} = 0.25$. A polydispersity of 0.3 for the particles has

been applied to the simulated curve. The agreement between experimental and calculated structure factors is satisfactory. The value of $\overline{n^{nano^2}}/\overline{n^{nano}}$ is moreover in good agreement with that determined by light scattering on the same solution. If the volume fraction of nanoparticles in the core were 0.5 as it is in the surfactant "supermicelles" [13,15], the core radius would be $R_C = 74$ Å and the corona thickness $h = R_H - R_C = 260$ Å. Such low values for the cores confirm the relatively weak structure peak found by SAXS around 0.16 Å$^{-1}$ (arrow in Fig. 4) and corresponding to the interparticle interactions.

In conclusion, we have demonstrated quantitatively the existence of hybrid organic-inorganic colloids made from densely packed nanoparticles and surrounded by a neutral polymer corona. The formation of the mixed colloids is controlled by the length of the neutral block, which role is crucial to prevent a macroscopic phase separation. The data shown in this letter suggest that the mechanism of their formation is electrostatic self-assembly.

**Acknowledgements** : We thank L. Novaki, K. Wong (Fisico-Quimica, Rhodia Brazil Ltda.) and L. Yang (Brookhaven National Laboratory) for their support in the SAXS experiments. M. Rawiso is acknowledged for having pointed out to us the derivation of the apparent molecular weight of polydisperse self-assembled colloids. This research was carried out in part at the National Synchrotron Light Source, Brookhaven National Laboratory, which is supported by the U.S. Department of Energy, Division of Materials Sciences and Division of Chemical Sciences, under Contract No. DE-AC02-98CH10886.

# Figure Captions

**Figure 1 :** Interparticle structure factors S(q,c) as determined by SAXS experiments on the aqueous suspensions of yttrium hydroxyacetate nanoparticles at the concentrations c = 0.8 (upper curve), 3, 6, 12 and 25.5 wt. % (lower curve). Inset : Guinier representation of the scattering intensity for low concentrated yttrium sols (c = 0.4 wt. %).

**Figure 2 :** Evolution of the hydrodynamic radius $R_H$ as a function of the mixing ratio X for solutions containing PANa(69)-*b*-PAM(840) copolymers and yttrium hydroxyacetate nanoparticles (c = 1 wt. %). At large X, the free nanoparticles tend to aggregate over time into sub-micrometric clusters.

**Figure 3 :** Zimm plot showing the evolution of the light scattering intensity measured for nanoparticle-polymer complexes as function of the wave-vector q. Straight lines are calculated according to Eq. 2 using for fitting parameters $R_G = 190 \pm 15$ Å, $M_{w,\text{app}} = 3.2 \times 10^6$ g mol$^{-1}$ and $A_2 = 10^{-6}$ cm$^{-3}$ g$^{-2}$.

**Figure 4 :** X-ray scattering intensities obtained for mixed nanoparticle-polymer solutions at c = 1 wt. % and X comprised between 0 and 10. Each curve has been shifted for sake of clarity. The upper curve for X = ∞ is the form factor of the yttrium-based particles (c < 0.4 wt. %). The arrow indicates the position of the correlation peak arising from particles interactions located in the cores [13,15].

**Figure 5 :** Comparison between the x-ray intensities obtained for the nanoparticles (form factor) and for the nanoparticle-polymer complexes at c = 1 wt. % and X = 0.2. Inset : Ratio between the two previous intensities, together with a fit (continuous line) using Eq. 3, $\overline{\mathrm{n}^{nano^2}}\big/\overline{\mathrm{n}^{nano}} = 39$ and $\overline{\mathrm{n}^{nano}} = 26$.

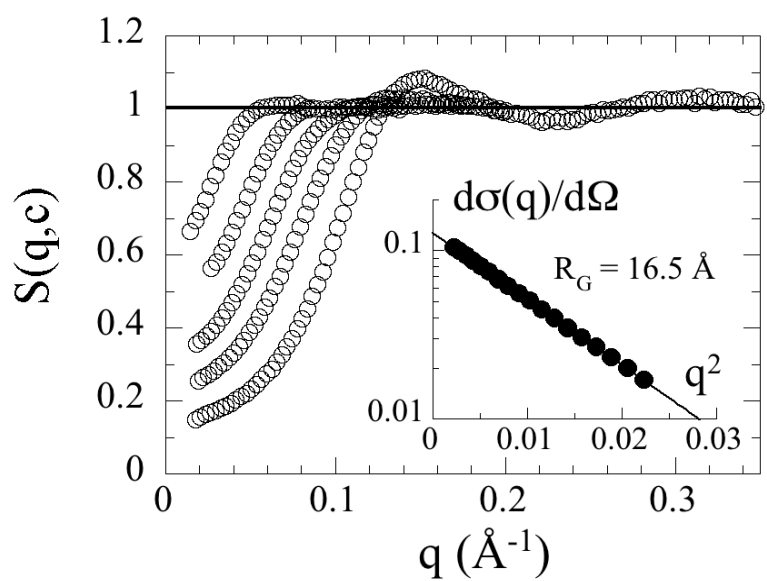

Figure 1

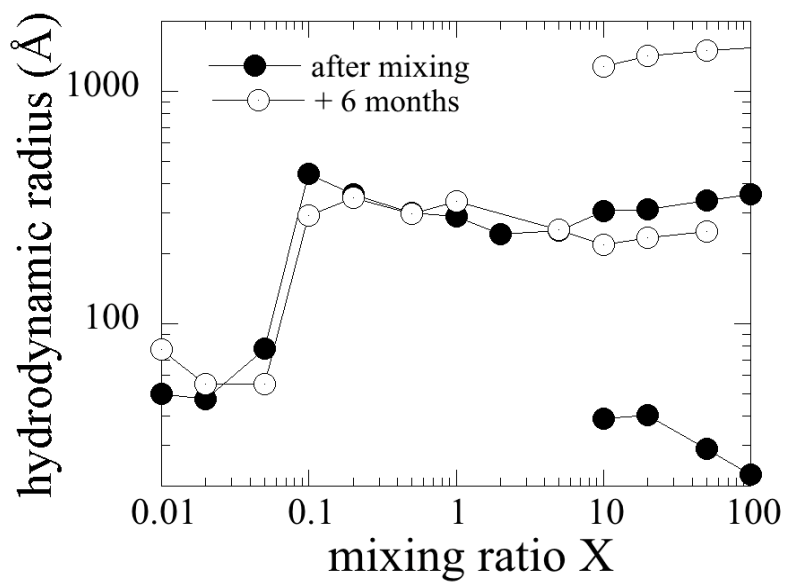

Figure 2

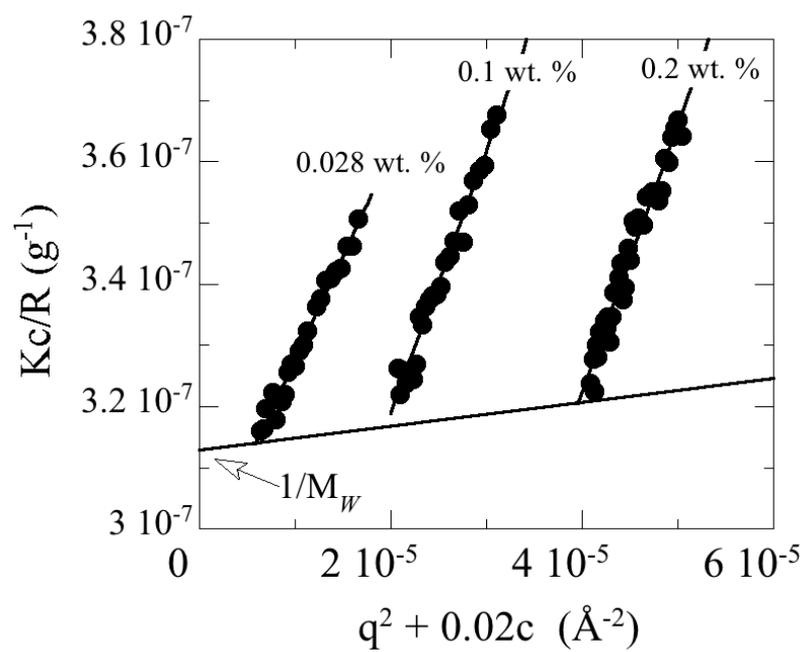

Figure 3

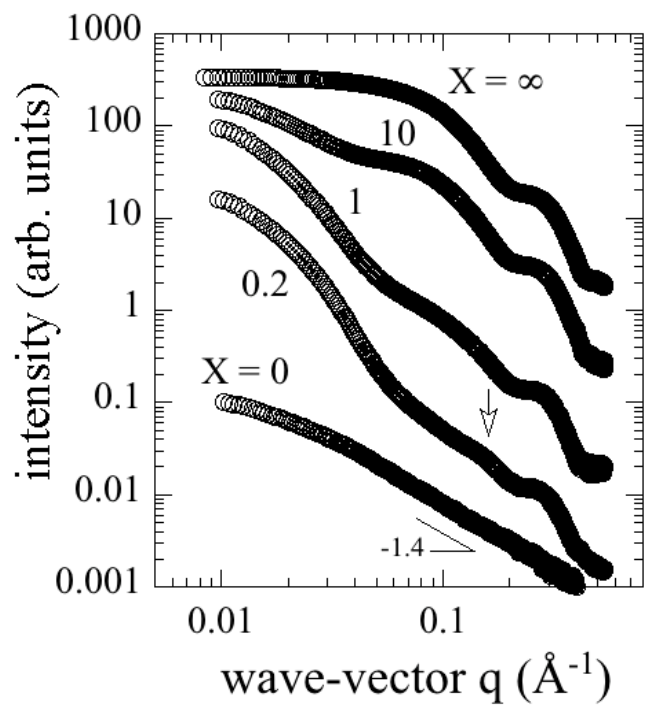

Figure 4

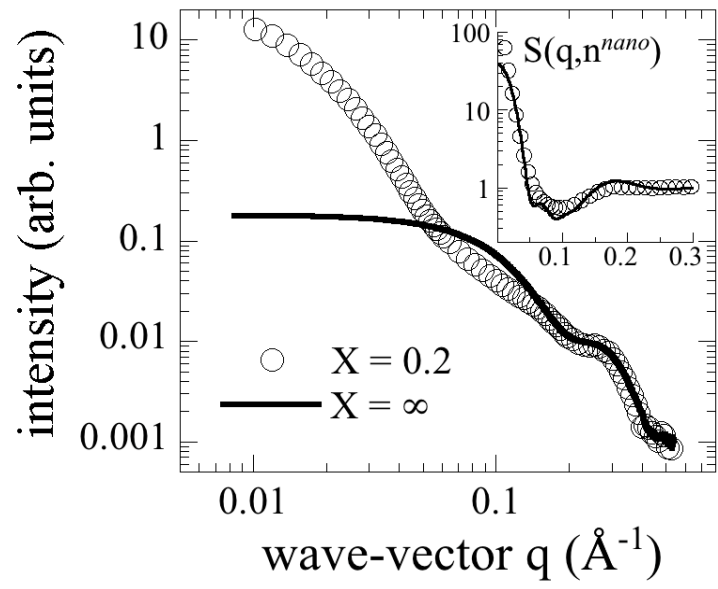

Figure 5